\documentclass[twocolumn,prl]{revtex4}
\usepackage{amsmath,amssymb}
\usepackage{pstricks}

\newcommand{\id}{\mathbf{1}}

\newcommand{\bl}{{\bf l}}

\newcommand{\brho}{\boldsymbol{\rho}}

\newcommand{\cM}{\mathcal{M}}

\begin{document}

\title{Domain walls, fusion rules and conformal field theory in the quantum Hall regime}

\author{Eddy Ardonne}
\affiliation{Nordita, Roslagstullsbacken 23, SE-106 91 Stockholm, Sweden}

\date{September 2, 2008}

\begin{abstract}
We provide a simple way to obtain the fusion rules associated with elementary quasi-holes
over quantum Hall wave functions, in terms of domain walls. The knowledge of
the fusion rules is helpful in the identification of the underlying conformal field
theory describing the wave functions. We obtain the fusion rules, and explicitly give a
conformal field theory description, for a two-parameter
family $(k,r)$ of wave functions. These include the Laughlin, Moore-Read and Read-Rezayi
states when $r=2$. The `gaffnian' wave function is the prototypical example for $r>2$, in
which case the conformal field theory is non-unitary.
\end{abstract}



\maketitle

Wave functions have played an instrumental role in the theoretical development of the
quantum Hall effect. The Laughlin wave function \cite{l83} predicted excitations with
fractional charge, which has been observed experimentally \cite{fraccharge}.
In the seminal work of Moore and Read \cite{mr91}, the connection between
conformal field theory (CFT) and wave functions was made, and a state in which
the excitations obey so-called non-abelian statistics was proposed.
There is ample numerical evidence (see \cite{numerics} for recent results)
that some states observed in the second Landau level harbor particles which
obey non-abelian statistics.  The possibility of non-abelian statistics has recently
spurred a tremendous amount of experimental effort \cite{experiment}, with
encouraging results, although a direct observation of non-abelian statistics is lacking so far.

By now, many of the proposed wave functions which are believed
to describe a quantum Hall state, such as the Read-Rezayi (RR) states \cite{rr99}, are written
as a CFT correlation function, or as a linear combination thereof, as is the case for
the Jain states \cite{hcjv} dominating the lowest Landau level.

The fundamental property underlying non-abelian sta\-tis\-tics is fusion, which
describes the possible outcomes of bringing two particles together.
We will denote the diffe\-rent types of particles (or quasi-holes) by $a$,
$b$, $c$, etc. The fusion of $a$ and $b$ is characterized by the non-negative
integers $N_{abc}$, which encode which particle types $c$ are
present in the fusion of the particles of type $a$ and $b$:
\begin{equation}
\label{fusion}
a \times b = \sum_{c} N_{abc} \, c \ .
\end{equation}
Particles for which there can appear more than one particle after fusion
with another particle are called non-abelian, because the Hilbert space
associated with several such particles is higher dimensional, opening up
the possibility of non-abelian braid statistics.

In this Letter, we will provide a simple, elegant way to obtain the fusion rules
associated to a two parameter family of $(k,r)$ wave functions without invoking CFT,
by examining the domain wall structure of the orbital occupation numbers. The fusion
rules obtained are those of $su(r)_k$. Building on this result, we provide an explicit
CFT description for the $(k,r)$ wave functions, which were investigated
by Haldane and Bernevig in terms of Jack polynomials \cite{bh08,bh08up}.
They reduce to the RR states \cite{rr99} in the case $r=2$. For $r>2$ (the
typical example being the gaffnian wave function, \cite{src07}), the CFT is
non-unitary, which implies, according to the arguments put forward in \cite{r08up},
that these wave functions describe  a critical phase, rather than a topological
state. We note that the described method to obtain the fusion rules is general, and can
help identifying the CFT, for instance for the wave functions considered in \cite{ww08up}.

{\em Orbital occupation numbers --}
The wave functions we consider here are characterized by the property that they
correspond to symmetric polynomials which do not vanish when
$k$ particles (which we will refer to as bosons) come together,
but have a zero of order $r$ when $k+1$ particles coincide.
It was shown in \cite{bh08,bh08up} that, for $r-1$ and $k+1$ relative prime,
these polynomials are Jack polynomials, with parameter
$\alpha=-\frac{k+1}{r-1}$ and labeled by a partition $\lambda$, which is related to
the orbital occupation number of the bosons (see below).
The orbital occupations are characterized by the rule that 
each set of $r$ neighboring orbitals contains exactly $k$ bosons, giving
$\tbinom{k+r-1}{k}$ different (bulk) patterns or sectors, which also naturally
arise in the thin torus limit \cite{thin-torus}.
To be explicit, we will give the sectors of the gaffnian \cite{src07}
with $(k,r)=(2,3)$ as an example (see also \cite{abk08} for the $r=2$ case).
In this case we have six sectors characterized by the following patterns of orbital
occupation numbers
$|l_0,l_1,l_2,\ldots\rangle$:
\begin{align*}
|200\, 200 \cdots\rangle && |020\, 020 \cdots\rangle && |002\, 002 \cdots\rangle \\ 
|110\, 110 \cdots\rangle && |101\, 101 \cdots\rangle && |011\, 011 \cdots\rangle 
\end{align*}
i.e. by the unit cells $(200)$, $(110)$ and their translations.

{\em Excitations and domain walls --}
One can consider quasi-holes by allowing configurations in which
$r$ neighboring orbitals contains less than $k$ bosons. The fundamental
quasi-holes, with smallest possible charge, correspond to configurations in which
there is only one set of $r$ neighboring
orbitals which contains $k-1$ bosons. It follows from the Su-Schrieffer counting argument \cite{ss81}
that these fundamental quasi-holes have charge $-e/r$, where $e$ is the charge of the
constituent bosons. The quasi-holes can be viewed as domain walls between different
sectors. The general structure is explained by using the gaffnian as an example.
In particular, we will look at the possible fundamental quasi-holes starting from
the $(110)$ sector:
\begin{align}
\label{fus1}
&| 1 1 0 \, 1{\bf10 \, 0}20 \, 020  \rangle
&| 1 1 0 \, 11{\bf0 \, 10}1 \, 101 \rangle
\end{align}
The boldface shows the location of the quasi-hole. Starting from the $(110)$ sector,
inserting a quasi-hole corresponds to a domain wall to either the $(020)$ or $(101)$
sector. These sectors are obtained from the $(110)$ sector by allowing one boson to hop one
orbital `to the right'.
This is the general structure: fundamental quasi-holes correspond to
domain walls between two sectors, where the unit cell of the resulting
sector is obtained from the initial one by hopping one boson one place to the right (assuming
periodic `boundary conditions' on the unit cell).
In general, the different sectors are in one-to-one correspondence with the
quasi-hole types. Starting from one sector, create a quasi-hole/particle pair, which
will have a new, different sector in between them. Move, say the quasi-hole around the torus.
After annihilating the quasi-hole/particle pair, the ground state will be the new sector.
Each different type of quasi-hole/particle pair will lead to a different ground state sector.

{\em Fusion rules --}
Because the domain walls discussed above correspond to the lowest charged quasi-hole,
and the sectors correspond to all the possible types of quasi-holes, we can interpret the domain walls
in terms of the fusion rules. Let us denote the lowest charged quasi-hole as a particle of type $a$.
If there is one fundamental domain wall connecting two sectors, say
$b$ and $c$, we interpret this by saying that sector $c$ is present in the fusion of $b$ with
the elementary quasi-hole $a$, i.e.  $N_{abc}=1$.
The possible fundamental domain walls completely specify the fusion rules of the particle type
$a$. In the quantum Hall case, we can obtain all types of particles by repeated fusion of
this fundamental quasi-hole.
This implies that all the fusion rules can be obtained from the fusion rules of $a$ by
associativity.
In general, it might not be obvious which sector corresponds
to the fundamental quasi-hole, but for the $(k,r)$ wave functions,
we can explicitly identify the fusion rules.

Before describing this general result, we first note that for $(k,r)=(2,3)$, we can identify the
six sectors in terms of the $su(3)_2$ representations (see below):
$(200)=\id$, $(110)={\bf 3}$,
$(101)=\overline{\bf 3}$, $(020)={\bf 6}$, $(011)={\bf 8}$ and $(002)=\overline{\bf 6}$,
where the last two numbers in each unit cell correspond to the $su(3)$ dynkin labels.
We interpret the two domain walls in \eqref{fus1} as the fusion rule
${\bf 3}\times {\bf 3} = \overline{\bf 3} + {\bf 6}$.

{\em Identification with $su(r)_k$ --}
To identify the fusion rules we obtained for the $(k,r)$ states above, we will map the
grounds state patterns to the labels of the irreducible representations of the affine
Lie algebra $su(r)_k$ (see \cite{dms} for an introduction).
The irreducible representations of $su(r)_k$ can be labeled by $r$
non-negative integers $(l_0;l_1,\ldots,l_{r-1})$, such that $\lambda = \sum_{i=1}^{r-1} l_i \omega_i$
is an $su(r)$ representation ($\omega_i$ are the fundamental weights), and $l_0$ is fixed by
$\sum_{i=0}^{r-1} l_i = k$. If this results in $l_0<0$, $\lambda$ does not correspond to an irreducible
representation of $su(r)_k$. This establishes a one-to-one correspondence between the
particle types of the $(k,r)$ states and the representations of $su(r)_k$. 

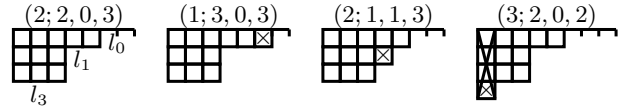
\begin{figure}[t]
\begin{center}
\psset{unit=.46mm,linewidth=.4mm,dimen=middle}
\begin{pspicture}(0,5)(40,25)
\multips(0,5)(5,0){3}{\psframe(0,0)(5,5)}
\multips(0,10)(5,0){3}{\psframe(0,0)(5,5)}
\multips(0,15)(5,0){5}{\psframe(0,0)(5,5)}
\psline(0,20)(35,20)
\psline(30,20)(30,18)
\psline(35,20)(35,18)
\rput(7.5,1){$l_3$}
\rput(20,11){$l_1$}
\rput(30,16){$l_0$}
\rput(17.5,24){$(2;2,0,3)$}
\end{pspicture}
\begin{pspicture}(0,5)(40,25)
\multips(0,5)(5,0){3}{\psframe(0,0)(5,5)}
\multips(0,10)(5,0){3}{\psframe(0,0)(5,5)}
\multips(0,15)(5,0){5}{\psframe(0,0)(5,5)}
\psline(0,20)(35,20)
\psline(30,20)(30,18)
\psline(35,20)(35,18)
\rput(27.5,17.5){$\times$}
\psframe(25,15)(30,20)
\rput(17.5,24){$(1;3,0,3)$}
\end{pspicture}
\begin{pspicture}(0,5)(40,25)
\multips(0,5)(5,0){3}{\psframe(0,0)(5,5)}
\multips(0,10)(5,0){3}{\psframe(0,0)(5,5)}
\multips(0,15)(5,0){5}{\psframe(0,0)(5,5)}
\psline(0,20)(35,20)
\psline(30,20)(30,18)
\psline(35,20)(35,18)
\rput(17.5,12.5){$\times$}
\psframe(15,10)(20,15)
\rput(17.5,24){$(2;1,1,3)$}
\end{pspicture}
\begin{pspicture}(0,5)(45,25)
\multips(0,5)(5,0){3}{\psframe(0,0)(5,5)}
\multips(0,10)(5,0){3}{\psframe(0,0)(5,5)}
\multips(0,15)(5,0){5}{\psframe(0,0)(5,5)}
\psline(0,20)(40,20)
\psline(30,20)(30,18)
\psline(35,20)(35,18)
\psline(40,20)(40,18)
\rput(2.5,2.5){$\times$}
\psframe(0,0)(5,5)
\rput(20,24){$(3;2,0,2)$}
\psline(0,0)(5,20)
\psline(0,20)(5,0)
\end{pspicture}
\caption{%
Leftmost diagram: the `Young diagram' associated to the $(2,2,0,3)$ representation of
$su(4)_7$. Remaing diagrams: the Young diagrams obtained after fusing $(2;2,0,3)$ with
$\omega_1=(6;1,0,0)$.
The crosses indicate the positions where the box of $\omega_1$ was added and
columns of height four are removed.}
\label{young}
\end{center}
\end{figure}

To obtain the fusion rules, we will make use of the Littlewood-Richardson (LR) rule \cite{dms}
for tensor products of $su(r)$ representations, which is stated in terms of the associated Young
diagrams.
A Young diagram is a set of rows of boxes, which are weakly decreasing in length. In general, the
$j$'th row has length $\sum_{i=j}^{r-1} l_i$. For our purposes, it will be useful to add a line of length
$\sum_{i=0}^{r-1} l_i$ to the top of the diagram, see figure \ref{young} for an $su(4)_7$ example.

We will only consider tensor products of arbitrary representations with one of the fundamental
weights $\lambda=\omega_i$, whose Young diagram is a single column of $i$ boxes.
We can obtain the associated fusion rules of $su(r)_k$ from the rule that in the fusion rules,
diagrams whose top row contains $k+1$ boxes are absent, as follows from the Kac-Walton
formula \cite{w90} relating tensor and fusion products. 

The LR rule specifies in which ways the boxes of the Young diagram of (in our case) $\omega_i$ 
can be added to the arbitrary representation $\lambda$, to obtain the representations in the
tensor product. The $i$ boxes of $\omega_i$ have to be added in such a way that the resulting
diagram is a Young-diagram (i.e. the length of the rows do not increase from top to bottom), and
no two boxes can be placed in the same row. It is allowed to place a box under the left-most column.
If this generates a column of height $r$, the whole column is to be removed, and $l_0$ adjusted
if necessary. To obtain the $su(r)_k$ fusion rules, we discard all
resulting diagrams whose top row contains $k+1$ boxes.

\begin{figure}[t]
\psset{unit=.46mm,linewidth=.4mm,dimen=middle}
\begin{pspicture}(0,5)(40,25)
\multips(0,5)(5,0){3}{\psframe(0,0)(5,5)}
\multips(0,10)(5,0){3}{\psframe(0,0)(5,5)}
\multips(0,15)(5,0){5}{\psframe(0,0)(5,5)}
\psline(0,20)(35,20)
\psline(30,20)(30,18)
\psline(35,20)(35,18)
\psframe(15,10)(20,15)
\psframe(25,15)(30,20)
\rput(17.5,12.5){$\times$}
\rput(27.5,17.5){$\times$}
\rput(17.5,24){$(1;2,1,3)$}
\end{pspicture}
\begin{pspicture}(0,5)(45,25)
\multips(0,5)(5,0){3}{\psframe(0,0)(5,5)}
\multips(0,10)(5,0){3}{\psframe(0,0)(5,5)}
\multips(0,15)(5,0){5}{\psframe(0,0)(5,5)}
\psline(0,20)(40,20)
\psline(30,20)(30,18)
\psline(35,20)(35,18)
\psline(40,20)(40,18)
\psframe(25,15)(30,20)
\psframe(0,0)(5,5)
\psline(0,0)(5,20)
\psline(5,0)(0,20)
\rput(27.5,17.5){$\times$}
\rput(2.5,2.5){$\times$}
\rput(20,24){$(2;3,0,2)$}
\end{pspicture}
\begin{pspicture}(0,5)(40,25)
\multips(0,5)(5,0){3}{\psframe(0,0)(5,5)}
\multips(0,10)(5,0){3}{\psframe(0,0)(5,5)}
\multips(0,15)(5,0){5}{\psframe(0,0)(5,5)}
\psline(0,20)(35,20)
\psline(30,20)(30,18)
\psline(35,20)(35,18)
\psframe(15,10)(20,15)
\psframe(15,5)(20,10)
\rput(17.5,12.5){$\times$}
\rput(17.5,7.5){$\times$}
\rput(17.5,24){$(2;1,0,4)$}
\end{pspicture}
\begin{pspicture}(0,5)(45,25)
\multips(0,5)(5,0){3}{\psframe(0,0)(5,5)}
\multips(0,10)(5,0){3}{\psframe(0,0)(5,5)}
\multips(0,15)(5,0){5}{\psframe(0,0)(5,5)}
\psline(0,20)(40,20)
\psline(30,20)(30,18)
\psline(35,20)(35,18)
\psline(40,20)(40,18)
\psline(0,0)(5,20)
\psline(5,0)(0,20)
\psframe(15,10)(20,15)
\psframe(0,0)(5,5)
\rput(17.5,12.5){$\times$}
\rput(2.5,2.5){$\times$}
\rput(20,24){$(3;1,1,2)$}
\end{pspicture}
\caption{The four possible fusion outcomes of fusing $(2;2,0,3)$ with $\omega_2$. The
crosses denote the positions where the boxes were added. Columns of height four are
removed.}
\label{omega2}
\end{figure}
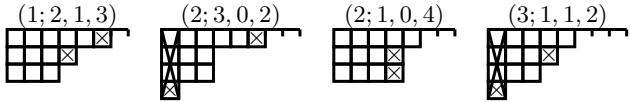

In figure \ref{young}, we give the Young diagram corresponding to $\lambda=(2;2,0,3)$, as well as
the diagrams resulting from fusion with $\omega_1=(6;1,0,0)$. The crosses denote the position
where the box of $\omega_1$ was added to the diagram of $\lambda$.
It is not hard to convince oneself that the LR rules presented above lead to the picture
that fusing an arbitrary representation $(l_0;l_1,\ldots,l_{r-1})$ with $\omega_1$ gives
maximally $r$ representations, characterized by
$l_i \rightarrow l_i-1$ and $l_{i+1}\rightarrow l_{i+1}+1$, for fixed  $i$, such that $l_i>0$ 
($l_0$ and $l_r$ are identified). These rules are exactly the ones we
found from the domain wall structure of the $(k,r)$ wave functions, which shows that the fusion
rules associated with the $(k,r)$ wave functions are the $su(r)_k$ fusion rules!
We can go one step further, and identify the fusion with an arbitrary $\omega_i$ in terms of domain
walls. In figure \ref{omega2}, we give the four possible fusion outcomes when one fuses $(2;2,0,3)$
with $\omega_2$.
We find that one can interpret fusing with $\omega_i$ in terms of the occupation numbers as
follows. The states in the fusion of a general representation are obtained by hopping $i$ bosons
one place to the right, with the constraint that from each position, one can only hop one boson
(which can be the one just hopped to that position). In terms of the
domain walls, this precisely corresponds to the situation in which there are $i$ strings of $r$
neighboring orbitals which have a deficit of one boson. A deficit of more than one boson in
a string of $r$ neighboring orbitals is
not allowed, because in the LR rule, this would correspond to placing two boxes in the same row.
We clarify this by using the gaffnian as an example, we find the following
`double' domain walls starting from the $(110)$ sector
\begin{align}
\label{fus2}
&| 1 1 {\bf 0 \, 100} \, 200 \, 200 \rangle 
&| 1 {\bf 1 0 \, 01}1 \, 011 \, 011 \rangle
\end{align}
In both cases, there are two strings of $3$ orbitals which have a deficit of one boson,
as indicated by the boldface. 
Like \eqref{fus1}, we can interpret \eqref{fus2} in terms of an $su(3)_2$ fusion rule,
namely ${\bf 3}\times \overline{\bf 3} = \id + {\bf 8}$.
 
 {\em CFT-construction --} Having identified the fusion rules, we will continue by giving an
explicit conformal field theory description of the $(k,r)$ wave functions. This construction
reduces to the known results for the gaffnian \cite{src07}, and
corroborates the results obtained from the study of Jack polynomials \cite{fjm02,bh08up}.
To get started, we will start by splitting off the $u(1)$-charge part of the theory, and consider the
remainder, containing the non-abelian structure. We will reinsert the charge part again in the end. 
We are after a two-parameter $(k,r)$ family of CFTs which for $r=2$ reduces to the
$Z_k$ parafermion CFT, describing the Read-Rezayi states \cite{rr99}.

As was anticipated in \cite{fjm02}, the CFTs needed are the minimal series
$(k+1,k+r)$ related to the $W_k$ algebra \cite{bs93}, which for $k=2$ is the
Virasoro algebra \cite{bpz84}. Here, we will explicitly give the operators creating the particles and
quasi-holes, and argue that they have the right properties to generate the
$(k,r)$ wave functions. To do this, we write the minimal models in terms of the coset \cite{mw90}
\begin{equation}
\frac{su(k)_1\times su(k)_{-\alpha-k}}{su(k)_{1-\alpha-k}} \ , \qquad \alpha=-\frac{k+1}{r-1}
\ . 
\end{equation}
We note that even though $\alpha$ is fractional for $r>2$, these cosets are well defined, but
non-unitary. These models are special cases of a more general set of minimal models
$\cM_k(p,p')$, (where $p$ and $p'$ are co-prime),  which reduce to the Virasoro minimal
models for $k=2$. In our case, we have $p=k+1$ and $p'=k+r$, and the central charge
given by $c=\frac{r(k-1)}{k+r}(1-k(r-2))$.
For $r=2$, the resulting coset is $su(k)_1\times su(k)_1/su(k)_2$, which
indeed corresponds to the $Z_k$ parafermions CFT.

We will refer to \cite{fl88,mw90} to obtain the field-content of the $\cM_{k}(k+1,k+r)$ models.
As usual, the coset fields carry labels of the constituent algebras.
In the case at hand, one can restrict oneself (by making use of field identifications \cite{mw90})
to the labels of $su(k)_r$. Thus, we write the fields as $\Phi_\bl$, where $\bl$ is vector of
$k-1$ non-negative integers whose sum does not exceed $r$.
The number of fields in this theory is given by $\binom{k+r-1}{r}$, and the fusion rules are identical
to the fusion rules of $su(k)_r$. We will show later that if one includes the charge sector, one
indeed obtains the correct $su(r)_k$ fusion rules for the full theory, in agreement with the domain
wall picture. The scaling dimensions of the fields $\Phi_\bl$, are given by \cite{sy88}
\begin{equation}
h_{\bl} =
\frac{k+1}{2(k+r)} \bl\cdot A_{k-1}^{-1}\cdot\bl -
\frac{r-1}{2(k+r)} \bl\cdot A_{k-1}^{-1}\cdot (2\brho) \ , 
\end{equation}
where $(A_{k-1}^{-1})_{i,j} = \min(i,j)-ij/k$ are the elements of the inverse Cartan matrix of $su(k)$, and
$\brho = (1,1,\ldots,1)$.

To establish that these models can be used to obtain the $(k,r)$ states, we will identify a class
of fields within these theories, which can be used as the creation operators for the bosons and
quasi-holes. The first set reduces to the $Z_k$ parafermion fields $\psi_i$ when $r=2$. These fields
$\psi_i^{(r)} = \Phi_{r \omega_i}$ have the same fusion rules as the $\psi_i$'s, namely 
$\psi^{(r)}_i \times \psi^{(r)}_j = \psi^{(r)}_{i+j\bmod k}$. Their scaling dimension is
$\Delta_{\psi_i} = \frac{r}{2}\frac{i(k-i)}{k}$. By making use of the same operator product expansion
arguments as those presented in \cite{rr99,gr00} (see also \cite{bh08up}),
one can show that the conformal correlator of $N$ operators
$\psi_1^{(r)} e^{i \phi \sqrt{r/k}}(z)$ (of dimension $r/2$), and a suitable background charge,
gives rise to the lowest degree symmetric polynomial which does not vanish when $k$
particles are brought at the same location, but vanishes with power $r$ when
$k+1$ particle positions coincide.
This shows that the $\cM_{k}(k+1,k+r)$ minimal models can be used to describe
the $(k,r)$ wave functions.

One can also identify the generalization of the $Z_k$ `spin-field' operators, namely
$\sigma_{i}^{(r)} = \Phi_{\omega_{i}}$, which have
scaling dimensions $\Delta_{\sigma_i} = \frac{i(k-i)(1-k(r-2))}{2k(k+r)}$.
When these fields are combined with the appropriate
vertex operator, they can be thought of as quasi-hole operators:
$V_{qh} (w) = \sigma_1^{(r)} e^{i \phi/\sqrt{rk}}(w)$.
One can show that these are the quasi-holes with the smallest possible charge, such that
the wave functions for the bosons and quasi-holes are analytic in the boson coordinates. 
It is in fact these operators which generate, upon fusion, all the sectors of the $su(r)_k$ states,
which are in one-to-one correspondence to the sectors of the $(k,r)$ states. 

For $r>2$, the lowest scaling dimension in the $\cM_k(k+1,k+r)$ model is negative, implying
that the model is non-unitary. For $k+1$ and $r-1$ co-prime, the lowest scaling dimension is
$h_{\rm min} = -\frac{kr(k-1)(r-2)}{24(k+r)}$.
From this, we find that the effective central charge, $c_{\rm eff} = c - 24 h_{\rm min}$
\cite{isz86}, is $c_{\rm eff} = (k-1)r/(k+r)$, which agrees with the conjecture put forward in \cite{bh08up}.

We will now argue that if one combines the $\cM_{k}(k+1,k+r)$ theory with the
$u(1)_{rk}$ chiral boson describing the charge,
one obtains the fusion rules of $su(r)_k$. This is a consequence of
rank level duality \cite{abi90}. In particular, the modular S-matrix of $su(r)_k$ can be written
in terms of the modular S-matrices of $su(k)_r$ and $u(1)_{rk}$ \cite{dms}, which relates the
fusion rules of $su(k)_r$ and $su(r)_k$.
We will not prove this in full generality, but rather demonstrate this
explicitly by considering the example of the gaffnian, which is described by the
$\cM_2(3,5)$ theory (which, using the notation of \cite{src07}, contains the fields
$\id, \sigma, \varphi, \psi$, with dimensions $0,-1/20,1/5,3/4$
and which obey $su(2)_3$ fusion rules). This theory is to
be combined with the chiral boson $u(1)_6$.
The six particle sectors of the full theory describing the gaffnian can be obtained by first
constructing the boson creation operator (which corresponds to the identity sector,
and is explicitly given by $\psi e^{3 i \phi/\sqrt{6}}$), and the smallest
charged quasi-hole $\sigma e^{i \phi/\sqrt{6}}$.
By subsequently fusing the quasi-hole, one obtains all the sectors of the
theory, which we give in table \ref{gaffnian}, where we also stated the corresponding $su(3)_2$
labels of the fields, the charge and scaling dimensions.
One can convince oneself that the six sectors of the gaffnian indeed satisfy the $su(3)_2$ fusion
rules.

\begin{table}[t]
\begin{tabular}{c||c|c|c|c|c|c}
qh & $\id$ & $\sigma e^{i \varphi/\sqrt{6}}$ &
$\phi e^{2i \varphi/\sqrt{6}}$ & $e^{2 i \varphi/\sqrt{6}}$ &
$\sigma e^{3i \varphi /\sqrt{6}}$ & $e^{4i \varphi /\sqrt{6}}$ \\
\hline
sector & $(200)$ & $(110)$ & $(101)$ & $(020)$ & $(011)$ & $(002)$ \\
$su(3)_2$ & $\id$ & ${\bf 3}$ &
$\overline{\bf 3}$ & ${\bf 6}$ & 
${\bf 8}$ & $\overline{\bf 6}$ \\
charge & $0$ & $\frac{1}{3}$ & $\frac{2}{3}$ & $\frac{2}{3}$ & 1 & $\frac{4}{3}$ \\
$h_{\rm qh}$ & $0$ & $\frac{1}{30}$ & $\frac{8}{15}$ & $\frac{1}{3}$ & $\frac{7}{10}$ & $\frac{4}{3}$ 
\end{tabular}
\caption{The sectors of the $v=\frac{2}{3}$ gaffnian wave function.}
\label{gaffnian}
\end{table}

We note that the rank-level duality also occurs in the Read-Rezayi states. The
$k+1$ fields in the full theory obey $su(2)_k$ fusion rules, but the $\frac{1}{2}k (k+1)$ fields of the 
associated $Z_k$ parafermion theory obey $su(k)_2$ fusion rules.

In conclusion, we presented a simple picture of the fusion-rules of non-abelian states in
terms of the occupation numbers. Elementary domain walls between regular patterns correspond
to the fusion rules of elementary quasi-holes. Having identified the fusion rules, we presented an
explicit CFT construction (following the conjecture in \cite{fjm02}, and \cite{bh08up}) of the
wave functions, based on
$\cM_{k}(k+1,k+r)$ `minimal' models, which are representations of the $W_{k}$-algebra.
The fusion rules are indeed equal to the fusion rules obtained from the domain walls.

Although these $su(r)_k$ wave functions for $r>2$ have a non-unitary conformal field theory
description, it is possible to construct wave functions associated to
manifestly unitary CFT's based on $su(r)_k$, such as the Ardonne-Schoutens
states \cite{as99}.
It would be interesting to investigate the relation between these quantum Hall states,
and (critical) wave functions, such as the gaffnian.
Another important question which needs further investigation is how the non-unitartiy 
for $r>2$ manifests itself in quantities calculable directly from the wave function, such
as edge correlations \cite{bh08up}.

{\em Acknowledgments --}
Stimulating discussions with B.A. Bernevig, H. Hansson,
A. Karlhede, N. Read, S. Simon, J. Slingerland and Z. Wang 
are gratefully acknowledged.

\end{document}